\documentclass{article}

\usepackage{hyperref}
\usepackage{stmaryrd}
\usepackage{amssymb}
\usepackage{graphicx}
\usepackage{tikz}
\usepackage{amsmath}
\usepackage{subcaption}
\usepackage{booktabs}
\usepackage{caption}
\usepackage{todonotes}
\usepackage{multirow}
\usepackage{amsfonts}
\usepackage{rotating}
\usepackage{url}
\usepackage{color}
\usepackage{xcolor}
\usepackage{array}
\usepackage{mathtools}
\usepackage{comment}
\usepackage{dblfloatfix}
\usepackage{enumitem}
\usepackage{algorithm}
\usepackage{algpseudocode}
\usepackage[margin=1.1in]{geometry}
\usepackage{multirow}
\usepackage{amsthm}
\usepackage{amsmath}
\usepackage{algorithmicx}
\usepackage[official]{eurosym}
\usepackage{pgfplots}
\usepackage{tablefootnote}
\usepackage{chessboard} 
\usepackage{xskak}      
\usepackage[affil-it]{authblk}  
\usetikzlibrary{shapes.geometric}
\usepackage[nameinlink]{cleveref}
\usepackage{adjustbox}
\usetikzlibrary{calc,arrows.meta, patterns}
\usetikzlibrary{tikzmark,decorations.pathreplacing,calc}
\usepackage{natbib}

\newtheorem{defi}{Definition}
\graphicspath{{Graphics/}}
\newcolumntype{L}[1]{>{\raggedright\arraybackslash}p{#1}}

\newcommand{\algmark}[1]{%
  \tikzmarknode[inner sep=0pt,outer sep=0pt]{#1}{\strut}%
}

\usepackage{microtype}
\newtheorem{lem}[defi]{Proposition}

\crefname{lem}{Prop.}{Props.}
\Crefname{lem}{Proposition}{Propositions}
\crefformat{lem}{%
  #2\textit{\small Prop.}~#1#3%
}
\crefmultiformat{lem}
  {#2\textit{\small Props.}~#1#3}
  {, #2#1#3}
  {, #2#1#3}
  {, and #2#1#3}

\date{\today}
\begin{document}
\title{Approximation guarantees for Hub Covering Problems}


\author[1]{Florian Jaehn}
\author[2]{Niklas Jost\thanks{Corresponding author: \texttt{niklas.jost@tu-dortmund.de}}}

\affil[1]{Institute for Management Science and Operations Research, Helmut Schmidt University, 22008 Hamburg, Germany}
\affil[2]{Chair of Material Handling and Warehousing, TU Dortmund University, 44227 Dortmund, Germany}

\maketitle

\begin{abstract}
\noindent
Hub Covering Problems are a subclass of Hub Location Problems. The objective is to select a set of hubs that enable paths between given origin-destination delivery tasks, while minimizing the total setup cost of the hubs. Two constraints must be satisfied: each path must include one or two hubs, and depending on the problem variant, the total path length or the length of the longest edge must not exceed prescribed limits. Such problems arise in a variety of practical applications, including urban planning, cargo delivery systems, airline networks, telecommunication networks, and e-mobility.

\noindent
Although Hub Covering Problems are especially important from a practical point of view, no systematic study of their approximability exists. In particular, it remains open whether some variants can be reduced to others in polynomial time, and no approximation bound is known. We close this gap by establishing a hierarchy among these problems, demonstrating that certain variants are indeed special cases of others. For four variants, we give a polynomial-time approximation algorithm, and for the other eight variants, we show that no polynomial-time algorithm can decide whether a solution exists unless $P=NP$.
\end{abstract}

	
	
	

\section{Introduction}

Hub Location Problems (HLPs) are integrated location and routing problems in which a set of \emph{hub nodes} must be selected to serve as intermediate points of paths between origin and destination nodes, referred to as \emph{branches}. Special cases of HLPs are Hub Covering Problems (HCPs), where a coverage radius restricts either the total path distance between two branches, the distance of any used edge, or the distance between a branch and its designated hub, depending on the problem variant.
This framework combines two essential aspects: (i) the consolidation and transfer of demand at hubs, which yield economies of scale, and (ii) coverage constraints, which ensure acceptable service levels. 

HCPs arise in any kind of logistic networks, such as distribution networks or communication networks, where coverage must satisfy legal, physical, or technical constraints within a combined location–routing framework. A real-world coverage constraint arises from working-hour regulations, where the distance of a driver’s tour is legally limited, as considered by \cite{goel2009vehicle}. Similarly, air transportation imposes range limits because aircraft can store only a finite amount of fuel; \cite{oktal2013hub} demonstrate that such range constraints are crucial when designing hub and spoke networks for aircraft. Moreover, \cite{esmizadeh2021cold} study hub networks that must preserve a cold chain: if a tour exceeds the coverage limit, the dry ice melts, and the temperature constraint is violated.

Devices such as access points, repeaters, routers, or transceivers transmit signals within a limited radius, naturally giving rise to coverage constraints. Meanwhile, traffic between sources and sinks uses a path through hub nodes, thereby reproducing the hub location structure (see \cite{farahani2013hub}). Recently, \cite{bollapragada2023hub} modeled the design of wireline broadband access networks as an HLP, ensuring that every user remains within a feasible signal range while aggregating traffic through the hub network. Similarly, \cite{funke2015placement} address the placement of electric-vehicle charging stations so that every demand location lies within a feasible battery range of at least one hub, with intermediate recharging enabling longer trips. These examples illustrate the dual coverage-plus-path structure.

Comprehensive surveys of HLPs are provided by \cite{alumur2008network} and \cite{farahani2013hub}. Coverage-based hub formulations were first introduced by \cite{campbell1994integer}, whose variants form the foundation of the present study. Afterwards, \cite{kara2003single} developed formulations and linearizations for HCPs, while \cite{wagner2008model} improved these formulations.
Approximation algorithms exist only for HLP variants without coverage constraints, such as those proposed by \cite{benedito2019approximation}, \cite{jost2023approximation}, and \cite{jost2025approximation}, which reduce HLPs to the Facility Location Problem. This reduction does not extend to HCPs due to their strict coverage limits. This raises the question which approximation bounds can be achieved by polynomial-time HCP algorithms and which algorithmic techniques may yield them.

Approximation results are well established for classical covering problems closely related to HCPs. For the set covering problem, \cite{chvatal1979greedy} proposed a greedy $O(\log n)$ (using standard asymptotic notation) approximation algorithm, which \cite{feige1998threshold} proved to be the best possible under $\mathrm{P}\neq\mathrm{NP}$. For the $p$-center problem, \cite{hochbaum1985best} established a tight factor-$2$ approximation (unless $\mathrm{P}=\mathrm{NP}$). Extensions incorporating capacities or reliability constraints have been studied by \cite{daskin2011network}. Although these models lack origin-destination relations through hubs, we will show that the result of \cite{feige1998threshold} can be transferred to HCPs.


The remainder of the paper is organized as follows: Section~\ref{chap pd} formally defines the HCPs. Section~\ref{chap p} presents approximation and hardness results. Section~\ref{chap:c} discusses implications and directions for future research.

\section{Preliminaries}\label{chap pd}
\subsection*{Problem Description}
We study twelve HCP variants following the definition of \cite{campbell1994integer}. We first describe the least restrictive one and derive the remaining ones by adding further constraints. We are given a weighted and undirected graph $G=(\mathcal{B}\cup \mathcal{H},E,d)$ in which the node set $\mathcal{B}\cup \mathcal{H}$ consists of \emph{branches} $\mathcal{B}$ and \emph{potential hubs} $\mathcal{H}$. Every \emph{edge} $ \{i,j\}\in E$ has a distance $d_{ij}=d_{ji}\geq 0$. For each potential hub $h\in\mathcal H$, a setup cost $c_h\geq 0$ is given. There is a set $\mathcal W \subseteq \mathcal B \times \mathcal B$ and each element $(b,b')\in \mathcal W$ defines an \emph{origin-destination (OD) delivery task}. For notational convenience, we assume symmetry in the delivery tasks such that $(b,b')\in\mathcal{W} \Rightarrow (b',b)\in\mathcal{W}$. Without loss of generality, we assume that for each branch $b\in \mathcal B$, there is at least one OD delivery task $(b,b')\in \mathcal W$. Finally, we are given a \textit{covering distance threshold} $\phi\in \mathbb{N}$, and a maximum number of hubs $m\in \mathbb{N}$. The input data for each HCP considered in this paper is summarized in the following Table \ref{tab:notation2}. 

\begin{table}[htbp]
\centering
\caption{Parameters}
\label{tab:notation2}
\begin{tabular}{lll}
\toprule
Symbol & Domain & Meaning \\
\midrule
$G$ & $G=(\mathcal{B}\cup \mathcal{H},E,d)$ & Graph \\
$\mathcal B$ &  - & Set of Branches \\
$\mathcal H$ &  - & Set of potential Hubs \\
$E$ & -         & Edge set \\
$d_{ij}$ &  $\mathbb{N}_0$ & Distance between nodes $i$ and $j$ \\
$c_h$ &  $\mathbb{N}_0$ & Setup cost for potential hub $h\in \mathcal H$ \\
$\mathcal{W}$ &  $\mathcal{W} \subseteq \mathcal{B}\times\mathcal{B}$ & Set of origin-destination delivery tasks \\
$\phi$ &  $\mathbb{N}$ & Coverage distance threshold \\
$m$ &  $\mathbb{N}$ & Upper bound on the number of open hubs \\
\bottomrule
\end{tabular}
\end{table}

For an OD delivery task $(b,b')\in \mathcal W$, a \emph{path} $p_{b,b'}$ is either a sequence of three vertices $(b,h,b')$ with $h\in \mathcal H$, $\{b,h\}, \{h,b'\}\in E$ or a sequence of four vertices $(b,h,h',b')$ with $h,h'\in \mathcal H$, $\{b,h\}, \{h',b'\}\in E$. Note that an edge $\{h,h'\}$ does not have to exist for the least restrictive variant. Without loss of generality, we assume $\{h,h\}\in E$ for all $h\in\mathcal H$ and $d_{hh}=0$ so that we can restrict ourselves to paths with four vertices in the following for every variant.

The problem is to determine a path for each OD delivery task. Thus, a \emph{solution} is a set $P$ of paths such that $P$ contains a path $p_{b,b'}$ for each OD delivery task $(b,b')\in\mathcal{W}$. If a potential hub $h\in \mathcal H$ is part of some path $p\in P$ (by abuse of notation, we write $h\in p$), we say that hub $h$ is \emph{open}. Thus, for a given solution $P$, the set of open hubs is $H(P):= \{h\in\mathcal H| \exists p\in P: h\in p\}$.

The objective is to minimize the sum of setup costs of open hubs, i.e., \
$\min \sum_{h \in H(P)} c_h$
is to be minimized.

A solution $P$ is \emph{feasible}, if the following constraint is satisfied:
\begin{description}
 \item[Branch to Hub Constraint (BH):] On every path, the branch-to-hub edges have at most distance $\phi$, i.e.\ for each $(b,h,h',b') \in P$:  $d_{bh}\leq \phi$ and $d_{b'h'}\leq \phi$ must hold.
\end{description}

This problem is denoted as multi allocation (MA), branch to hub constrained (BH) variant without cardinality constraints (noCC) (MA-BH-noCC).

\subsection*{Problem Variants}

In addition to BH, we introduce two further covering constraints, each of which makes the previous redundant. These constraints are based on the variants proposed by \cite{campbell1994integer}.

\begin{description}
 \item[Edge Constraint (E):] The distance of every edge on a path must not exceed $\phi$, i.e.\ for each  $(b,h,h',b') \in P$: $\{h,h'\}\in E$, $d_{bh}\leq \phi$, $d_{hh'}\leq \phi$ and $d_{h'b'}\leq \phi$ must hold.
 \item[Total Path Constraint (TP):] The sum of the edge distances of a path must not exceed $\phi$, i.e.\ for each $(b,h,h',b') \in P$: $\{h,h'\}\in E$ and $d_{bh}+ d_{hh'}+ d_{h'b'}\leq \phi$ must hold.
\end{description}

Besides one of the three covering constraints, two additional constraints are part of some problem variants. The first constraint affects the case if a branch $b$ is part of more than one OD delivery task. In general, the corresponding paths may connect $b$ with different hubs for each OD delivery task. In the following constraint, all paths containing branch $b$ must contain the very same edge connecting $b$ to a hub.

\begin{description}
 \item[Single Allocation (SA):] For any two OD delivery tasks $(b,b'),(b,b'')\in \mathcal{W}$ there must be a hub $h$ such that the corresponding paths $p_{b,b'},p_{b,b''}\in P$ both contain edge $\{b,h\}$.
\end{description}
If the SA option is not part of the problem setting, we refer to the corresponding variants as \emph{multi allocation (MA)} variants, as in our original problem. The second additional constraint limits the number of open hubs.
\begin{description}
 \item[Cardinality constraint (CC):] The number of open hubs must not be greater than $m$, i.e.\ $|H(P)|\leq m$.
\end{description}

If CC is not part of the problem setting, we refer to the corresponding variants as \emph{without cardinality constraint (noCC)} variants, as in our original problem. From a modeling perspective, it is sufficient to set $m\geq \mathcal |\mathcal{H}|$ as this is always fulfilled.

A full factorial design of the three covering constraint attributes \{BH, E, TP\} ($\blacksquare$), the two allocation type attributes \{SA, MA\} ($\triangleright$), and the two cardinality constraint attributes \{noCC, CC\} ($\bullet$) leads to twelve problem variants, which are summarized below and in Figure \ref{fig:visual}.

\begin{figure}

\begin{tikzpicture}[
    >=Latex,
    line join=round,
    line cap=round,
    font=\small
]

\def\W{1.85}      
\def\H{1.15}      
\def\ox{0.30}     
\def\oy{0.28}     
\def\Xsep{2.25}
\def\Ysep{1.65}

\coordinate (D) at (2,3.5);

\pgfmathsetmacro{\Xmax}{2*\Xsep+\W+\ox}
\pgfmathsetmacro{\Ymin}{-\Ysep}
\pgfmathsetmacro{\Ymax}{\H+\oy}

\colorlet{frontNoCC}{black!6}
\colorlet{topNoCC}{black!12}
\colorlet{sideNoCC}{black!18}

\colorlet{frontCC}{black!18}
\colorlet{topCC}{black!26}
\colorlet{sideCC}{black!34}

\tikzset{
    pattMA/.style={pattern=north east lines, pattern color=black!65},
    pattSA/.style={pattern=dots, pattern color=black!75},
    proj/.style={black!45, densely dashed},
    slab/.style={black!30},
    axis/.style={thick,-},
    varlabel/.style={
        font=\scriptsize,
        align=center,
        inner sep=1.2pt,
        rounded corners=0.7pt,
        fill=white,
        fill opacity=0.82,
        text opacity=1
    }
}


\newcommand{\BHSwatch}{%
\tikz[baseline=-0.6ex,scale=0.22]{
    \fill[black!8]  (0,0) rectangle (1.2,0.7);
    \fill[black!15] (0,0.7) -- (0.25,0.95) -- (1.45,0.95) -- (1.2,0.7) -- cycle;
    \fill[black!22] (1.2,0) -- (1.45,0.25) -- (1.45,0.95) -- (1.2,0.7) -- cycle;
    \draw (0,0) rectangle (1.2,0.7);
    \draw (0,0.7) -- (0.25,0.95) -- (1.45,0.95) -- (1.2,0.7);
    \draw (1.2,0) -- (1.45,0.25) -- (1.45,0.95);
}}

\newcommand{\ESwatch}{%
\tikz[baseline=-0.6ex,scale=0.22]{
    \fill[black!10] (0.7,0.5) circle (0.42);
    \draw (0.7,0.5) circle (0.42);
    \draw[black!55] (0.28,0.53) arc[start angle=180,end angle=360,x radius=0.42,y radius=0.14];
    \draw[black!35,dashed] (1.12,0.53) arc[start angle=0,end angle=180,x radius=0.42,y radius=0.14];
}}

\newcommand{\TPSwatch}{%
\tikz[baseline=-0.6ex,scale=0.22]{
    \fill[black!8]  (0.2,0) rectangle (1.0,0.8);
    \fill[black!15] (0.2,0.8) -- (0.45,1.05) -- (1.25,1.05) -- (1.0,0.8) -- cycle;
    \fill[black!22] (1.0,0) -- (1.25,0.25) -- (1.25,1.05) -- (1.0,0.8) -- cycle;
    \draw (0.2,0) rectangle (1.0,0.8);
    \draw (0.2,0.8) -- (0.45,1.05) -- (1.25,1.05) -- (1.0,0.8);
    \draw (1.0,0) -- (1.25,0.25) -- (1.25,1.05);
}}
\newcommand{\PatternSwatch}[1]{%
\tikz[baseline=-0.6ex]{
    \fill[white] (0,0) rectangle (0.34,0.22);
    \fill[#1] (0,0) rectangle (0.34,0.22);
    \draw (0,0) rectangle (0.34,0.22);
}}

\newcommand{\GraySwatch}[1]{%
\tikz[baseline=-0.6ex]{
    \fill[#1] (0,0) rectangle (0.34,0.22);
    \draw (0,0) rectangle (0.34,0.22);
}}


\newcommand{\BHshape}[7]{%
\begin{scope}[shift={(#1,#2)}]
    \def\bw{1.85}
    \def\bh{0.88}

    \fill[#3] (0,0) rectangle (\bw,\bh);
    \fill[#4] (0,\bh) -- (\ox,\bh+\oy) -- (\bw+\ox,\bh+\oy) -- (\bw,\bh) -- cycle;
    \fill[#5] (\bw,0) -- (\bw+\ox,\oy) -- (\bw+\ox,\bh+\oy) -- (\bw,\bh) -- cycle;

    \fill[#6] (0,0) rectangle (\bw,\bh);

    \draw (0,0) rectangle (\bw,\bh);
    \draw (0,\bh) -- (\ox,\bh+\oy) -- (\bw+\ox,\bh+\oy) -- (\bw,\bh);
    \draw (\bw,0) -- (\bw+\ox,\oy) -- (\bw+\ox,\bh+\oy);

    \node[varlabel] at (.5*\bw,.5*\bh) {#7};
\end{scope}
}

\newcommand{\Eshape}[5]{%
\begin{scope}[shift={(#1,#2)}]
    \coordinate (C) at (0.93,0.58);
    \def\r{0.67}

    \fill[#3] (C) circle[radius=\r];

    \begin{scope}
        \clip (C) circle[radius=\r];
        \fill[#4] ($(C)+(-\r,-\r)$) rectangle ($(C)+(\r,\r)$);
    \end{scope}

    \draw (C) circle[radius=\r];

    \draw[black!55] ($(C)+(-0.50,0.06)$)
        arc[start angle=180,end angle=360,x radius=0.50,y radius=0.18];
    \draw[black!35,dashed] ($(C)+(0.50,0.06)$)
        arc[start angle=0,end angle=180,x radius=0.50,y radius=0.18];

    \draw[black!35] ($(C)+(-0.22,0.48)$)
        arc[start angle=150,end angle=20,x radius=0.26,y radius=0.13];

    \node[varlabel] at (C) {#5};
\end{scope}
}

\newcommand{\TPshape}[7]{%
\begin{scope}[shift={(#1,#2)}]
    \def\bw{1.15}
    \def\bh{1.15}
    \def\xs{0.35}

    \fill[#3] (\xs,0) rectangle (\xs+\bw,\bh);
    \fill[#4] (\xs,\bh) -- (\xs+\ox,\bh+\oy) -- (\xs+\bw+\ox,\bh+\oy) -- (\xs+\bw,\bh) -- cycle;
    \fill[#5] (\xs+\bw,0) -- (\xs+\bw+\ox,\oy) -- (\xs+\bw+\ox,\bh+\oy) -- (\xs+\bw,\bh) -- cycle;

    \fill[#6] (\xs,0) rectangle (\xs+\bw,\bh);

    \draw (\xs,0) rectangle (\xs+\bw,\bh);
    \draw (\xs,\bh) -- (\xs+\ox,\bh+\oy) -- (\xs+\bw+\ox,\bh+\oy) -- (\xs+\bw,\bh);
    \draw (\xs+\bw,0) -- (\xs+\bw+\ox,\oy) -- (\xs+\bw+\ox,\bh+\oy);

    \node[varlabel] at (\xs+0.5*\bw,0.5*\bh) {#7};
\end{scope}
}

\foreach \j/\alloc/\patt in {0/MA/pattMA,1/SA/pattSA}{
    \pgfmathsetmacro{\Y}{-\j*\Ysep}

    \BHshape{0}{\Y}{frontNoCC}{topNoCC}{sideNoCC}{\patt}{BH-\alloc\\-noCC}
    \Eshape{\Xsep}{\Y}{frontNoCC}{\patt}{E-\alloc\\-noCC}
    \TPshape{2*\Xsep}{\Y}{frontNoCC}{topNoCC}{sideNoCC}{\patt}{TP-\alloc\\-noCC}
}

\begin{scope}[shift={(D)}]
\foreach \j/\alloc/\patt in {0/MA/pattMA,1/SA/pattSA}{
    \pgfmathsetmacro{\Y}{-\j*\Ysep}

    \BHshape{0}{\Y}{frontCC}{topCC}{sideCC}{\patt}{BH-\alloc\\-CC}
    \Eshape{\Xsep}{\Y}{frontCC}{\patt}{E-\alloc\\-CC}
    \TPshape{2*\Xsep}{\Y}{frontCC}{topCC}{sideCC}{\patt}{TP-\alloc\\-CC}
}
\end{scope}

\coordinate (Fsw) at (-0.18,\Ymin-0.12);
\coordinate (Fse) at (\Xmax+0.18,\Ymin-0.12);
\coordinate (Fnw) at (-0.18,\Ymax+0.12);
\coordinate (Fne) at (\Xmax+0.18,\Ymax+0.12);

\coordinate (Bsw) at ($(Fsw)+(D)$);
\coordinate (Bse) at ($(Fse)+(D)$);
\coordinate (Bnw) at ($(Fnw)+(D)$);
\coordinate (Bne) at ($(Fne)+(D)$);

\draw[proj] (Fsw) -- (Bsw);
\draw[proj] (Fse) -- (Bse);
\draw[proj] (Fnw) -- (Bnw);
\draw[proj] (Fne) -- (Bne);

\draw[slab] (Fsw) rectangle (Fne);
\draw[slab] (Bsw) rectangle (Bne);

\coordinate (O)  at (-0.45,\Ymin-0.45);   
\coordinate (Ox) at (\Xmax+0.70,\Ymin-0.45);
\coordinate (Oy) at (-0.45,\Ymax-0.3);
\coordinate (Oz) at ($(O)+(3.1,5.40)$);

\fill (O) circle (1pt);

\draw[axis] (O) -- (Ox);

\foreach \i/\cov/\icon in {0/BH/\BHSwatch,1/E/\ESwatch,2/TP/\TPSwatch}{
    \pgfmathsetmacro{\XC}{\i*\Xsep + 0.5*\W}
    \draw[black!50] (\XC,\Ymin-0.53) -- (\XC,\Ymin-0.37);
    \node[below=5pt,align=center] at (\XC,\Ymin-0.45)
        {\icon\\[-1pt]\scriptsize \cov};
}

\node[anchor=north, align=center]
    at ($(O)!0.5!(Ox)+(0,-1.05)$)
    {\(\blacksquare\) covering constraint};

\draw[axis] (O) -- (Oy);

\draw[black!50] (-0.52,0.5*\H) -- (-0.38,0.5*\H);
\draw[black!50] (-0.52,\Ymin+0.5*\H) -- (-0.38,\Ymin+0.5*\H);

\node[left=4pt,align=right] at (-0.45,0.5*\H)
    {\PatternSwatch{pattMA}\;MA};
\node[left=4pt,align=right] at (-0.45,\Ymin+0.5*\H)
    {\PatternSwatch{pattSA}\;SA};

\node[rotate=90, anchor=south, align=center]
    at ($(O)!0.52!(Oy)+(-1.35,0)$)
    {\(\triangleright\) allocation};

\coordinate (Ov) at ($(Ox)+(Oz)-(O)$);

\coordinate (Cno) at ($(Ox)!0.10!(Ov)$);
\coordinate (Ccc) at ($(Ox)!0.70!(Ov)$);
\coordinate (OvShort) at ($(Ox)!0.80!(Ov)$);

\draw[axis] (Ox) -- (OvShort);

\path
    ($(Cno)+(1.75,-0.10)$) --
    ($(Ccc)+(1.75,-0.10)$)
    node[
        midway,
        sloped,
        anchor=center,
        align=center
    ]
    {\(\bullet\) cardinality constraint};

\draw[black!55]
    ($(Cno)+(-0.11,0.06)$) -- ($(Cno)+(0.11,-0.06)$);
\draw[black!55]
    ($(Ccc)+(-0.11,0.06)$) -- ($(Ccc)+(0.11,-0.06)$);

\node[anchor=west, align=left]
    at ($(Cno)+(0.18,0.10)$)
    {\GraySwatch{frontNoCC}\;{\scriptsize noCC}};

\node[anchor=west, align=left]
    at ($(Ccc)+(0.18,0.10)$)
    {{\scriptsize CC}\;\GraySwatch{frontCC}};

\end{tikzpicture}

 \caption{The twelve problems under consideration}
    \label{fig:visual}
\end{figure}

\begin{itemize}
    \item[$\blacksquare$] The distance of each branch-to-hub edge on a path must not exceed~$\phi$. (\textbf{BH})
    \item[$\blacksquare$] The distance of each edge on a path must not exceed~$\phi$. (\textbf{E})
    \item[$\blacksquare$] The sum of all edge distances on a path must not exceed~$\phi$. (\textbf{TP})
    \item[$\triangleright$] Each path can be considered individually. (\textbf{MA})
    \item[$\triangleright$] Each branch must be connected to the same hub for every path. (\textbf{SA})
    \item There is no restriction in the number of open hubs. (\textbf{noCC})
    \item The number of open hubs is restricted by $m$. (\textbf{CC})
\end{itemize}

\subsection*{Model}

In the following, we represent all twelve problem variants using a mixed integer linear program.
We introduce the following decision variables:

\begin{itemize}
    \item $Y_h \in \{0,1\}$: indicates whether hub $h\in\mathcal{H}$ is opened, i.e., $Y_h=1$ iff $h\in p$ for some $p\in P$.
    \item $X_{bhh'b'} \in \{0,1\}$: indicates whether path $(b,h,h',b')$ is chosen to cover delivery task $(b,b') \in \mathcal{W}$, i.e., $X_{bhh'b'} =1$ iff $(b,h,h',b')\in P$.
    \item $Z_{bh} \in \{0,1\}$: indicates in the SA setting whether edge $\{b,h\}$ is contained in all solution paths for OD delivery tasks $(b,b') \in \mathcal{W}$.
    
\end{itemize}

\begin{align}
    \min \quad & \sum_{h\in \mathcal{H}} c_h Y_h && & \text{All} \label{eq:obj}\\
    \noalign{\vskip 10pt}
    \text{s.t.} \quad& \sum_{h,h'\in \mathcal{H}} X_{bhh'b'} \ge 1 && \forall (b,b')\in \mathcal{W} & \text{All}\label{co:4}\\
    &X_{bhh'b'}=X_{b'h'hb}& &\forall b,b'\in \mathcal{B},\, h,h'\in \mathcal{H} & \text{All}\label{co:25} \\
    & 2\cdot X_{bhh'b'} \le Y_h+Y_{h'} && \forall b,b'\in \mathcal{B},\, h,h'\in \mathcal{H}& \text{All} \label{co:5}\\
    \noalign{\vskip 10pt}
     &    d_{bh} X_{bhh'b'} \leq \phi      && \forall (b,b')\in \mathcal{W}, \; \forall h,h'\in \mathcal{H}& \text{All}\label{co:8} \\
    &    d_{hh'} X_{bhh'b'} \leq \phi      && \forall (b,b')\in \mathcal{W}, \; \forall h,h'\in \mathcal{H}& \text{E, TP}\label{co:10} \\
    &     (d_{bh} +  d_{hh'} + d_{h'b'})\, X_{bhh'b'} \leq \phi      && \forall (b,b')\in \mathcal{W}, \; \forall h,h'\in \mathcal{H}& \text{TP} \label{co:7}\\
  \noalign{\vskip 10pt}
    &X_{bhh'b'}\leq Z_{bh} &&\forall b,b'\in \mathcal{B},\, h,h'\in \mathcal{H} & \text{SA}\label{co:32}\\
    &X_{bhh'b'}\geq Z_{bh}+Z_{b'h'}-1& & \forall b,b'\in \mathcal{B},\, h,h'\in \mathcal{H} & \text{SA}\label{co:42}\\
    & \sum_{h \in \mathcal{H}} Z_{bh} = 1 && \forall b \in \mathcal{B} & \text{SA}\label{co:3}\\
  \noalign{\vskip 10pt}
    &\sum_{h\in \mathcal{H}} Y_h \leq m && & \text{CC}\label{co:11}\\   
    \noalign{\vskip 10pt}
    & Y_h \in \{0,1\} && \forall h\in \mathcal{H} & \text{All} \label{co:12}\\
    & X_{bhh'b'} \in \{0,1\}&& \forall b,b'\in \mathcal{B},\, h,h'\in \mathcal{H} & \text{All}\label{co:13}\\
    &Z_{bh}\in \{0,1\} && \forall b\in \mathcal{B},\, h\in \mathcal{H} & \label{co:14}\text{SA}
\end{align}

The objective~\eqref{eq:obj} minimizes total hub opening costs. 
The constraint formulation consists of five logical blocks.

The first block guarantees the feasibility of the hub configuration.
Constraint~\eqref{co:4} requires every OD delivery task to be covered by at least
one admissible path. Constraint~\eqref{co:25} ensures symmetry on the paths.
Constraint~\eqref{co:5} restricts admissible paths to opened hubs,
thereby coupling routing decisions with hub opening decisions.

The second block contains the distance-based covering constraints.
Constraint~\eqref{co:8} bounds the lengths of branch-to-hub edges and is
active in all variants.
Constraint~\eqref{co:10} additionally restricts hub-to-hub connections for the E and TP variant.
Constraint~\eqref{co:7} restricts the total path length in the TP variant. Note that constraint \eqref{co:7} implies constraints \eqref{co:8}-\eqref{co:10}. We nevertheless include them in the TP formulation to emphasize that TP is a more restrictive version than E or BH.

The third block enforces the single allocation structure. The linking between the single allocation variables and the path-based
representation ($X_{bhh'b'}=Z_{bh}\cdot Z_{b'h'}$) is linearized by Constraint \eqref{co:32} and \eqref{co:42}. In combination with the symmetry constraint~\eqref{co:25}, this linearization already implies the $X_{bhh'b'}\leq Z_{b'h'}$ condition. Constraint~\eqref{co:3} ensures that each branch is mapped to exactly one hub
in the single allocation setting.

The fourth block limits the number of open hubs if the CC is active.
Constraint~\eqref{co:11} enforces the cardinality constraint when present.

The final block of constraints specifies the domains of the decision variables.
Constraints~\eqref{co:12}–\eqref{co:14} impose integrality conditions.

Note that Constraints~\eqref{co:8} and~\eqref{co:10} can be ensured by a preprocessing step that deletes all edges \(e\) with distance \(d_e>\phi\). Moreover, Constraint~\eqref{co:13} can be relaxed to a continuous constraint in the SA variants, since \(Z\) is integral: if \(Z_{bh}=0\) or \(Z_{b'h'}=0\), then Constraints~\eqref{co:25} and~\eqref{co:32} force \(X_{bhh'b'}=0\); if \(Z_{bh}=Z_{b'h'}=1\), then Constraint~\eqref{co:42} forces \(X_{bhh'b'}=1\).

\subsection*{Robustness Under Metric and Discount Variations}

In classical hub location problems, it is common to assume a \emph{discount factor} 
$\alpha \in [0,1]$ applied to hub-to-hub distances in order to model economies of scale.
Moreover, instances are often assumed to be defined on complete metric graphs.

For the hardness and approximation results established in this paper,
neither the metric assumption nor the explicit inclusion of a discount factor is essential.

The discount factor can be absorbed into the input distances
by replacing each hub-to-hub distance $d_{hh'}$ with $\alpha d_{hh'}$.
Hence, explicitly modeling $\alpha$ does not change the combinatorial
structure of the problem.
All lower bounds established below remain valid for $\alpha=1$,
and therefore also hold for $\alpha \in [0,1]$.

We call an instance \emph{pseudo-metric} if there exists a value 
$\alpha \in (0,1]$ such that, after dividing every hub-to-hub distance
by $\alpha$, the resulting graph is complete and metric.
This notion reflects the standard modeling in hub location,
where hub-to-hub distances are discounted uniformly,
while allowing us to avoid explicitly carrying $\alpha$
in every formulation.

The reductions used in our hardness proofs rely only on threshold comparisons
of the form $d \le \phi$.
For the BH and E variants, feasibility depends solely on whether distances
are below or above the threshold~$\phi$.
Thus, any instance can be transformed into an equivalent instance on a complete metric graph
by replacing distances $d \leq \phi$ to $\phi$
and distances $d > \phi$ or previous non-existing edges to $2\cdot \phi$.
This transformation preserves feasibility and objective function values.

For the TP variant, the lower bound is obtained by a polynomial-time
reduction from the E variant to a complete pseudo-metric TP instance.
Consequently, the hardness result holds even under metric assumptions
and in the presence of a discount factor.

For clarity of exposition, we therefore omitted explicit metric assumptions
and the discount factor $\alpha$ in the formal problem statement.
All results remain valid if these modeling features are included.

\section{Results}\label{chap p}

Table~\ref{tab:variants} summarizes, for each problem variant, a lower and an upper bound on the achievable approximation ratio or claims that no approximation guarantee exists unless $P = NP$. We use standard asymptotic notation \(O(\cdot)\), \(\Omega(\cdot)\), and \(\Theta(\cdot)\) to denote upper, lower, and tight bounds on the worst-case approximation ratio.

\begin{table}[htbp]
\centering
\caption{Approximation guarantees of hub covering problem variants.}
\label{tab:variants}
\begin{adjustbox}{max width=\textwidth}
\begin{tabular}{c|cc|cc}
\toprule
 & \multicolumn{4}{c}{Allocation type} \\
\cmidrule(lr){2-5}
& \multicolumn{2}{c}{Single allocation (SA)} 
& \multicolumn{2}{c}{Multi allocation (MA)} \\
\cmidrule(lr){2-3} \cmidrule(lr){4-5}
Covering constraint & noCC & CC & noCC & CC \\
\midrule
Branch--hub (BH) 
& $\Theta(\log|\mathcal{B}|)$~\cref{lem142} & no approx.~\cref{lem142} & $\Theta(\log|\mathcal{B}|)$~\cref{lem42} & no approx.~\cref{lem42} \\
Edge-based (E) 
& no approx.~\cref{lem3} & no approx.~\cref{lem6} & $\Omega(\log|\mathcal{B}|)\land O(|\mathcal{B}|^2)$~\cref{lem6,lem4} & no approx.~\cref{lem6} \\
Total path (TP) 
& no approx.~\cref{lem4}  & no approx.~\cref{lem4} & $\Omega(\log|\mathcal{B}|) \land O(|\mathcal{B}|^2)$~\cref{lem4,lem2} & no approx.~\cref{lem4} \\
\bottomrule
\end{tabular}
\end{adjustbox}
\end{table}

These results, summarized in Table~\ref{tab:variants}, are obtained via reductions to and from the Set Cover Problem (\cref{lem42}), from the $n$-Queens Completion problem (\cref{lem3}), and via reductions between different variants of the problem itself (\cref{lem142,lem4,lem6}).
In addition, we construct an approximation algorithm (\cref{lem2}).

Since two of the following proofs rely on polynomial-time reductions from classical covering and placement problems, we first recall the Set Cover Problem and the $n$-Queens Completion Problem. Afterwards, we state and prove the propositions.

\paragraph{Set Cover Problem}

Let $U = \{ e_1, e_2, \dots, e_n \}$ be a finite set called the \emph{universe}, and let $\mathcal{S} = \{ S_1, S_2, \dots, S_m \}$ be a collection of subsets of $U$, i.e.\ $S_i \subseteq U$ for all $i \leq m$.
The goal of the Set Cover Problem (SCP) is to find a minimum-size subcollection $\mathcal{C} \subseteq \mathcal{S}$ such that every element of $U$ is contained in at least one set in $\mathcal{C}$.

Formally, introducing variables $x_i = 1$ if set $S_i$ is chosen and $0$ otherwise, the problem can be formulated as:

\begin{align}
    \min \quad & \sum_{i=1}^{m} x_i \label{SCP:obj}\\
    \text{s.t.} \quad 
        & \sum_{i : e_j \in S_i} x_i  \geq 1,
        && \forall j = 1,\dots,n, 
        &&  
        \label{SCP:1}\\
    & x_i \in \{0,1\},
        && \forall i = 1,\dots,m,
        && 
        \label{SCP:2}
\end{align}

The weighted version assigns each set $S_i$ a weight $w_i > 0$,
and seeks to minimize the total weight such that the objective is changed to $\min {\sum_{i=1}^{m} w_i x_i}$. We call the unweighted version \emph{uSCP} and the weighted one \emph{wSCP}. 

In the decision variant of the uSCP (\emph{duSCP}), an additional integer $p$ is given, and it must be decided if there is an uSCP solution with objective at most $p$. This version is NP-complete (see \cite{garey2002computers}).

\cite{lund1994hardness} showed that the uSCP has an approximation lower bound of 
$\Omega(\log |U|)$ unless $P = NP$. Furthermore, \cite{chvatal1979greedy} introduced a greedy algorithm achieving an 
$O(\log |U|)$ approximation ratio for the wSCP.

\paragraph{$n$-Queens Completion Problem}

In the \textit{$n$-Queens Problem}, a natural number $n$ is given. A feasible solution consists of an arrangement of $n$ queens on an $n\times n$ chessboard such that no pair of queens are in the same row, column, or diagonal. Formally, a solution consist of a set of $n$ tuples $\{(x_1,y_1),...(x_n,y_n)\}$ with $x_i,y_i\in \{1,...,n\}$ for all $i\in \{1,...,n\}$. Additionally, $x_i\neq x_j$, $y_i \neq y_j$ and $|x_i-x_j|\neq |y_i-y_j|$ for all pairs $i,j\in \{1,...,n\}$ with $i\neq j$. If and only if two queens are in the same row, column, or diagonal, we say that these queens are \emph{attacking} each other.

The \emph{$n$-Queens Completion Problem} is an extension with an additional given set $Q_0 \subseteq \{1,\dots,n\} \times \{1,\dots,n\}$ of pre-placed queens. The task is to extend $Q_0$ to a full placement of $n$ queens on the board such that no two queens are attacking each other. This problem is NP-complete (see ~\cite{gent2017complexity}).
Moreover, \cite{glock2022n} showed that there is always a valid $n$-Queens completion solution with $n>3$ and at most $\frac{n}{60}$ pre-placed queens. Therefore, we consider at least $\frac{n}{60}$ pre-placed queens such that the input of this problem is always in $O(n)$.

\begin{lem}\label{lem42}
Unless $P=NP$, the MA-BH-noCC variant is $\Theta(\log |\mathcal{B}|)$-approximable and there is no approximation guarantee for MA-BH-CC. These results hold even if all setup costs are $1$.
\end{lem}

\textbf{Proof:}
First, we show a polynomial-time reduction from MA-BH-noCC to the wSCP with $|U|=|\mathcal{B}|$,
preserving feasibility and objective function value.
As there is a polynomial-time $O(\log |U|)$ approximation algorithm for wSCP, this bound carries over to MA-BH-noCC.

Second, we construct a polynomial-time reduction from the uSCP to MA-BH-noCC and from the duSCP to MA-BH-CC, where the number of branches matches the number of elements in the universe $|\mathcal{B}|=|U|$.
This reduction preserves uniform costs, i.e., all hubs in the constructed MA-BH-noCC and MA-BH-CC instance have setup cost $c_h = 1$.
Since any uSCP approximation algorithm is in $\Omega(\log |U|)$ unless $P = NP$,
the $\Omega(\log |\mathcal{B}|)$ lower bound applies to MA-BH-noCC, even under uniform setup costs. Since no polynomial-time algorithm can decide for duSCP whether a solution exists (unless $P=NP$), this result carries over to MA-BH-CC. 

We defer the proof that the reductions in parts one and two preserve feasibility to a third part.

\paragraph{1. Reduction from MA-BH-noCC to wSCP}\

Given an MA-BH-noCC instance, we construct a wSCP instance as follows:

\begin{itemize}
    \item For each branch $b_j \in \mathcal{B}$, define an element $e_j$.
    \item For each hub $h_i \in \mathcal{H}$, define a set $S_i$ with weight $w_i = c_i$.
    \item Let $e_j \in S_i$ if and only if $\{b_j,h_i\}$ exists and  $d_{b_jh_i} \le \phi$. 
\end{itemize}

Figure~\ref{xxx} illustrates an MA-BH-noCC instance, while Figure~\ref{figurexxx} shows the corresponding wSCP instance obtained by this construction.

\begin{figure}[H]
\begin{center}
\begin{tikzpicture}[scale=1.5]
    \node[shape=circle,draw=blue,minimum size=25pt] (u1) at (-2,2) {$b_1$};
    \node[shape=circle,draw=blue,minimum size=25pt] (u2) at (-1,2) {$b_2$};
    \node[shape=circle,draw=blue,minimum size=25pt] (u3) at (1,2) {$b_3$};
    \node[shape=circle,draw=blue,minimum size=25pt] (u4) at (2,2) {$b_4$};
    \node[shape=circle,draw=orange,minimum size=25pt] (s1) at (-2,0) {$h_1$};
    \node[shape=circle,draw=green!70!black,minimum size=25pt] (s2) at (0,0) {$h_2$};
    \node[shape=circle,draw=purple,minimum size=25pt] (s3) at (2,0) {$h_3$};
    
    \draw (u1) -- (s1);
    \draw (u2) -- (s1);
    \draw (u2) -- (s2);
    \draw (u3) -- (s2);
    \draw (u3) -- (s3);
    \draw (u4) -- (s3);
    \draw (s1) -- (s2);
\end{tikzpicture}
\end{center}
\caption{MA-BH-noCC instance where different setup costs correspond to different colors of the hubs and edges have distances at most $\phi$.}
\label{xxx}

	\end{figure}

\begin{figure}[H]
\begin{center}
\begin{tikzpicture}
    
    \node[draw=orange!50, thick, ellipse, minimum width=6cm, minimum height=4cm] (H1) at (-4,0) {};
    \node[below=0.5cm, text width=2cm, align=center] at (H1.south) {\textbf{$S_1$}};

    \node[draw=green!70!black!50, thick, ellipse, minimum width=6cm, minimum height=4cm] (H2) at (0,0) {};
    \node[below=0.5cm, text width=2cm, align=center] at (H2.south) {\textbf{$S_2$}};

    \node[draw=purple!50, thick, ellipse, minimum width=6cm, minimum height=4cm] (H3) at (4,0) {};
    \node[below=0.5cm, text width=2cm, align=center] at (H3.south) {\textbf{$S_3$}};
    
    
    \node[draw=blue, circle] at (-4,0) {$e_1$}; 
    \node[draw=blue, circle] at (-2,0) {$e_2$}; 
    \node[draw=blue, circle] at (2,0) {$e_3$}; 
    \node[draw=blue, circle] at (4,0) {$e_4$}; 
\end{tikzpicture}
\end{center}
	\caption{wSCP instance obtained from the MA-BH-noCC instance in Figure \ref{figurexxx}. The weights are equal to the corresponding hub setup costs.}
\label{figurexxx}

	\end{figure}

Since we constructed a bijective mapping of hubs from $H$ to sets $S$, we can also map solutions bijectively, i.e. a solution for the wSCP with subcollection $\mathcal{C}=\{S_{\overline{1}},S_{\overline{2}},...,S_{\overline{l}}\}$ maps to a solution for MA-BH-noCC with open hubs $\{h_{\overline{1}},h_{\overline{2}},...,h_{\overline{l}}\}$ and vice versa. The construction can be carried out in polynomial time.

The objective function values are identical:
\[\underbrace{\sum_{i=1}^{m}  w_{i}x_{i}
~=~
\sum_{i=1}^{l}  w_{\overline{i}}x_{\overline{i}}}_{\text{wSCP objective}}
~=~ 
\underbrace{\sum_{i=1}^{l}  c_{\overline{i}}Y_{\overline{i}}~=~
\sum_{h\in\mathcal{H}}  c_{h}Y_{h}}_{\text{MA-BH-noCC objective}}
\]

To complete the reduction, we show that these solutions are either both feasible or infeasible. As this proof is the same as for the second reduction, we defer it to the third part.

\paragraph{2. Reduction from uSCP to MA-BH-noCC and duSCP to MA-BH-CC.}

Let an instance of the (decision variant of the) unweighted Set Cover Problem be given by a universe
$U = \{e_1,\dots,e_n\}$, a family of sets $\mathcal{S} = \{S_1,\dots,S_m\}$ and a maximal number of usable sets $p$ (for duSCP).
We construct an MA-BH-noCC (resp. MA-BH-CC) instance as follows:

\begin{itemize}
    \item For each element $e_j \in U$, create a branch $b_j \in \mathcal{B}$.
    \item For each set $S_i \in \mathcal{S}$, create a hub $h_i \in \mathcal{H}$.
    \item Set the hub opening cost to $c_{h_i} = 1$ for every hub.
    \item Set $\phi=1$.
    \item Create an edge $\{b_j,h_i\}$ if and only if $e_j \in S_i$ and set $d_{b_j h_i} =1$. 
    \item The set of OD delivery tasks is defined by
    \[
        \mathcal{W} = {\{(b_1,b_2),(b_2,b_1),(b_1,b_3),(b_3,b_1)},(b_1,b_4),\dots,(b_n,b_1)\}.
    \]
    \item If reducing from duSCP to MA-BH-CC, set the maximal number of open hubs $m$ to $m=p$.
\end{itemize}

Note that every branch is part of at least two OD delivery task.
As in the reduction in the first part, a solution for the (d)uSCP with subcollection $\mathcal{C}=\{S_{\overline{1}},S_{\overline{2}},...,S_{\overline{l}}\}$ bijectively maps to a solution for MA-BH-noCC (resp. MA-BH-CC) with open hubs $\{h_{\overline{1}},h_{\overline{2}},...,h_{\overline{l}}\}$ and vice versa. The construction can be carried out in polynomial time.

The objective function values for uSCP and the MA-BH-noCC are identical as:

\[
\underbrace{
\sum_{i=1}^{m} x_i
~=~
\sum_{i=1}^{l} x_{\overline{i}}
}_{\text{uSCP objective}}
~=~ l ~=~
\underbrace{
\sum_{i=1}^{l} Y_{\overline{i}}
~=~
\sum_{i=1}^{l} c_{\overline{i}}Y_{\overline{i}}
~=~
\sum_{h\in\mathcal{H}} c_hY_h
}_{\text{MA-BH-noCC objective}}
\]

To complete the proof, we show that the solutions for the uSCP and the MA-BH-noCC (resp. duSCP and MA-BH-CC) are either both infeasible or have the same objective function value.

\paragraph{3. Feasibility}

When the open hubs are fixed, feasible delivery paths can be computed greedily by testing any combination of open hubs for each delivery task.

We now show that if and only if $\mathcal{C}$ is an infeasible uSCP/wSCP/duSCP solution, there is no feasible MA-BH-noCC/MA-BH-CC solution with exactly $\{h_{\overline{1}},h_{\overline{2}},...,h_{\overline{l}}\}$ as open hubs.

If $\mathcal{C}$ is infeasible in duSCP as it uses more than $p$ sets, it is also infeasible for MA-BH-CC as it opens more than $m$ hubs.
Otherwise, if $\mathcal{C}$ is an infeasible SCP solution, this is because some element $e_j\in U$ is uncovered in the sets in $\mathcal{C}$, i.e.\ $e_j \not \in \bigcup_{S\in \mathcal{C}} S$.
By construction, branch $b_j$ appears in at least one OD delivery task $(b_j,b')$. If there is a feasible path $(b_j,h_i,h',b')$, we need edge $\{b_j,h_i\}$ to exist implying $e_j\in S_i$ and $h_i\in \{h_{\overline{1}},h_{\overline{2}},...,h_{\overline{l}}\}$ implying $S_i\in \{S_{\overline{1}},S_{\overline{2}},...,S_{\overline{l}}\}$. This contradicts $e_j \not \in \bigcup_{S\in \mathcal{C}} S$ and no feasible MA-BH-noCC (resp. MA-BH-CC) solution exists. 

Vice versa, if $\mathcal{C}$ is a feasible SCP solution, then fewer than $p$ hubs are open and for every element $e$ there is at least one set $S\in \mathcal{C}$ covering it, i.e.\ $e \in \bigcup_{S\in \mathcal{C}} S\quad \forall e\in U$. If the MA-BH-noCC (resp. MA-BH-CC) solution is infeasible there is an OD delivery task $(b_i,b_j)$ for which no path $(b_i,h_{i'},h_{j'},b_j)\quad \forall h_{i'},h_{j'}\in \{h_{\overline{1}},h_{\overline{2}},...,h_{\overline{l}}\}$ exists. As $e_i \in \bigcup_{S\in \mathcal{C}} S$ and $e_j \in \bigcup_{S\in \mathcal{C}} S$ there are some open hubs $h_{i'},h_{j'}$ such that the edges $\{b_i,h_{i'}\}$ and $\{b_j,h_{j'}\}$ exists with distances at most $\phi$. Then path $(b_i,h_{i'},h_{j'},b_j)$ is feasible for OD delivery task $(b_i,b_j)$ contradicting the assumption. \qed

\begin{lem}\label{lem142}
Unless $P=NP$, the SA-BH-noCC variant is $\Theta(\log |\mathcal{B}|)$-approximable and there is no approximation guarantee for SA-BH-CC.
\end{lem}

\textbf{Proof:}
In Proposition \ref{lem42}, we have shown the claim for the corresponding MA variants. We now show that SA-BH-noCC (resp. SA-BH-CC) has the same approximation guarantee as MA-BH-noCC (resp. MA-BH-CC). It is sufficient to show this for the covering constraint variant, as setting $m\ge |H|$ will give the claim for the noCC variant.

Every feasible SA-BH-CC solution is feasible for MA-BH-CC with the same objective function value. It remains to show that every feasible MA-BH-CC solution can be transformed into a feasible SA-BH-CC solution without increasing the objective function value.
We give a polynomial-time reduction from MA-BH-CC to SA-BH-CC.

Let a feasible MA-BH-CC solution be given. We construct a feasible SA-BH-CC solution for the same problem instance as follows:

\begin{itemize}
    \item Open the same hubs as in the MA variant. 
    \item For every branch $b_i$, consider an arbitrary path $(b_i,h_i,h',b')$ or $(b',h',h_i,b_i)$ in the MA solution. Set $Z_{b_i,h_i}=1$ and any $Z_{b_i,h_j}=0$ for all $h_j\neq h_i$ giving that all OD delivery paths including $b_i$ use $\{b_i,h_i\}$.
    \item Choose for every OD delivery task $(b_i,b_j)$ the delivery path $(b_i,h_i,h_j,b_j)$ with $Z_{b_i,h_i}=Z_{b_j,h_j}=1$.
\end{itemize}

As the SA and MA solutions use the same set of hubs, they have the same objective function value.

Assume, for contradiction, that this construction gives an infeasible SA solution, although the MA solution was feasible. By construction, the SA solution features the SA requirement. If a path $(b_i,h_i,h_j,b_j)$ is infeasible, either edge $\{b_i,h_i\}$ or edge $\{b_j,h_j\}$ is infeasible in any delivery path. However, every edge appearing in an SA delivery path also appears in at least one MA delivery path. This contradicts the assumption that the MA delivery paths are feasible while the SA delivery paths are not. \qed

\begin{lem}\label{lem3}
Deciding whether SA-E-noCC allows for a feasible solution is an NP-complete problem.
\end{lem}

\textbf{Proof:}
Membership in NP is immediate, since the feasibility of a given solution can be
verified in polynomial time. 
We prove the claim by a polynomial-time reduction from the
\emph{$n$-Queens Completion Problem}.

Let an instance of the $n$-Queens Completion Problem be given. 
We construct a corresponding SA-E-noCC instance as follows:
\begin{itemize}
    \item For each row $i \in \{1,\dots,n\}$, we create a branch $b_i$.
    \item For each square $(i,j)\in \{1,\dots,n\}^2$ of the board, we create a potential hub $h_{i,j}$.
    \item Set $\phi=1$.
    \item Create a hub-to-hub edge $\{h_{i,j},h_{k,l}\}$ if and only if $i\neq k$, $j\neq l$ and $|i-k|\neq |j-l|$ and set $d_{h_{i,j}h_{k,l}}=1$. 
    \item Create a branch-to-hub edge $\{b_{i},h_{k,l}\}$ if and only if $i= k$ and $(i,j)\not \in Q_0\ \forall j\neq l$ and set $d_{b_{i}h_{k,l}}=1$. 
    \item Create OD delivery tasks $\mathcal{W}=\mathcal{B}\times \mathcal{B}$ for every pair of branches.
\end{itemize}

Figure~\ref{fig:n-Queens-example} illustrates a small $3$-Queens completion
instance with one pre-placed queen.
Figure~\ref{figure} shows the corresponding SA-E-noCC instance obtained by this construction.

\begin{figure}[H]
    \centering

    \begin{minipage}[t]{0.34\textwidth}
        \centering
        \vspace{0pt}
        \begin{minipage}[c][6.2cm][c]{\linewidth}
            \centering
            \newchessgame
            \setchessboard{
                maxfield=c3,
                setpieces={qa2},
                showmover=false,
                labelbottomformat=\arabic{filelabel}
            }
            \chessboard[boardfontsize=40]
        \end{minipage}

        \caption{Example of a $3$-Queens Completion Problem with a queen on \((2,1)\).}
        \label{fig:n-Queens-example}
    \end{minipage}
    \hfill
    \begin{minipage}[t]{0.60\textwidth}
        \centering
        \vspace{0pt}
        \begin{minipage}[c][6.2cm][c]{\linewidth}
            \centering
            \begin{tikzpicture}[scale=0.78, transform shape]

                \node[shape=circle,draw=lightgray] (11) at (2,6) {\( h_{3,1} \)};
                \node[shape=circle,draw=lightgray] (12) at (4,6) {\( h_{3,2} \)};
                \node[shape=circle,draw=lightgray] (13) at (6,6) {\( h_{3,3} \)};
                \node[shape=circle,draw=lightgray] (21) at (2,4) {\( h_{2,1} \)};
                \node[shape=circle,draw=lightgray] (22) at (4,4) {\( h_{2,2} \)};
                \node[shape=circle,draw=lightgray] (23) at (6,4) {\( h_{2,3} \)};
                \node[shape=circle,draw=lightgray] (31) at (2,2) {\( h_{1,1} \)};
                \node[shape=circle,draw=lightgray] (32) at (4,2) {\( h_{1,2} \)};
                \node[shape=circle,draw=lightgray] (33) at (6,2) {\( h_{1,3} \)};
                \node[shape=circle,draw=blue,minimum size=33pt] (1) at (-2,6) {$b_3$};
                \node[shape=circle,draw=blue,minimum size=33pt] (2) at (-2,4) {$b_2$};
                \node[shape=circle,draw=blue,minimum size=33pt] (3) at (-2,2) {$b_1$};

                \draw[-, bend right,line width=0.4mm] (3) edge (31);
                \draw[-, bend right,line width=0.4mm] (3) edge (32);
                \draw[-, bend right,line width=0.4mm] (3) edge (33);

                \draw[-, bend left,line width=0.4mm] (1) edge (13);
                \draw[-, line width=0.4mm] (2) edge (21);
                \draw[-, bend left,line width=0.4mm] (1) edge (12);
                \draw[-, bend left,line width=0.4mm] (1) edge (11);

                \draw[-,line width=0.4mm] (11) edge (32);
                \draw[-,line width=0.4mm] (13) edge (32);
                \draw[-,line width=0.4mm] (12) edge (31);
                \draw[-,line width=0.4mm] (12) edge (33);
                \draw[-,line width=0.4mm] (21) edge (33);
                \draw[-,line width=0.4mm] (23) edge (31);
                \draw[-,line width=0.4mm] (21) edge (13);
                \draw[-,line width=0.4mm] (23) edge (11);

            \end{tikzpicture}
        \end{minipage}

        \caption{Converted construction from the example in Figure~\ref{fig:n-Queens-example}.}
        \label{figure}
    \end{minipage}

\end{figure}

Note that we depart from standard chess notation by numbering both rows and columns. This avoids using $b1$ to denote both a square on the board and the first branch.
The construction involves $O(n^2)$ hubs and $O(n^4)$ edges and can be carried out in
polynomial time.
In the following, we show that a solution for the $n$-Queens Completion Problem $\{(x_1,y_1),(x_2,y_2),...,(x_n,y_n)\}$ corresponds to a solution for SA-E-noCC with open hubs $\{h_{x_1,y_1},h_{x_2,y_2},...,h_{x_n,y_n}\}$ and vice versa. For any OD delivery task $(b_i,b_j)$ we consider the path $(b_i,h_{x_i,y_i},h_{x_j,y_j},b_j)$. Figure~\ref{figure2} highlights an invalid SA-E-noCC solution, corresponding to a
queen placement in which two queens attack each other, as shown in
Figure~\ref{fig:n-Queens-example2}. To complete the reduction, we show that these solutions are either both feasible or infeasible. This is done by showing that for every feasible $n$-Queens Completion Problem solution, the corresponding SA-E-noCC solution is also feasible and vice versa.

\begin{figure}[H]
    \centering

    \begin{minipage}[t]{0.58\textwidth}
        \centering
        \vspace{0pt}
        \begin{minipage}[c][6.2cm][c]{\linewidth}
            \centering
            \begin{tikzpicture}[scale=0.78, transform shape]

                \node[shape=circle,draw=lightgray] (11) at (2,6) {\( h_{3,1} \)};
                \node[shape=circle,draw=lightgray] (12) at (4,6) {\(h_{3,2} \)};
                \node[shape=circle,draw=red] (13) at (6,6) {\( h_{3,3} \)};
                \node[shape=circle,draw=red] (21) at (2,4) {\( h_{2,1} \)};
                \node[shape=circle,draw=lightgray] (22) at (4,4) {\( h_{2,2} \)};
                \node[shape=circle,draw=lightgray] (23) at (6,4) {\( h_{2,3} \)};
                \node[shape=circle,draw=lightgray] (31) at (2,2) {\( h_{1,1} \)};
                \node[shape=circle,draw=red] (32) at (4,2) {\( h_{1,2} \)};
                \node[shape=circle,draw=lightgray] (33) at (6,2) {\( h_{1,3} \)};
                \node[shape=circle,draw=blue,minimum size=33pt] (1) at (-2,6) {$b_3$};
                \node[shape=circle,draw=blue,minimum size=33pt] (2) at (-2,4) {$b_2$};
                \node[shape=circle,draw=blue,minimum size=33pt] (3) at (-2,2) {$b_1$};

                \draw[-, bend right,line width=0.4mm] (3) edge (31);
                \draw[-,red, bend right,line width=0.4mm] (3) edge (32);
                \draw[-, bend right,line width=0.4mm] (3) edge (33);

                \draw[-, bend left,line width=0.4mm] (1) edge (12);
                \draw[-, bend left,line width=0.4mm] (1) edge (11);
                
                \draw[-,red, bend left,line width=0.4mm] (1) edge (13);
                \draw[-,red, line width=0.4mm] (2) edge (21);

                \draw[-,line width=0.4mm] (11) edge (32);
                \draw[-,cyan,line width=0.4mm] (13) edge (32);
                \draw[-,line width=0.4mm] (12) edge (31);
                \draw[-,line width=0.4mm] (12) edge (33);
                \draw[-,line width=0.4mm] (21) edge (33);
                \draw[-,line width=0.4mm] (23) edge (31);
                \draw[-,cyan,line width=0.4mm] (21) edge (13);
                \draw[-,line width=0.4mm] (23) edge (11);

            \end{tikzpicture}
        \end{minipage}

        \caption{The red hubs \( h_{2,1},h_{1,2},h_{3,3} \) are open. This solution does not serve OD delivery task $(b_1,b_2)$ since edge \(\{h_{2,1},h_{1,2}\}\) does not exist.}
        \label{figure2}
    \end{minipage}
    \hfill
    \begin{minipage}[t]{0.36\textwidth}
        \centering
        \vspace{0pt}
        \begin{minipage}[c][6.2cm][c]{\linewidth}
            \centering
            \newchessgame
            \setchessboard{
                maxfield=c3,
                setpieces={qc3, qa2, qb1},
                showmover=false,
                labelbottomformat=\arabic{filelabel}
            }
            \chessboard[boardfontsize=40]
        \end{minipage}

        \caption{Corresponding $3$-queens completion solution to Figure~\ref{figure2}. Invalid since the queen on \((2,1)\) attacks the queen on \((1,2)\).}
        \label{fig:n-Queens-example2}
    \end{minipage}

\end{figure}

Let a feasible solution to the $n$-Queens Completion Problem be given. We show that the corresponding SA-E-noCC solution is feasible by checking the SA constraint and the existence of feasible paths for every OD delivery task.
In every feasible $n$-Queens Completion Problem solution, exactly one queen is placed in each row, giving $x_i\neq x_j\quad \forall i\neq j$. By construction, this implies for every branch $b$ the existence of exactly one open hub $h$ such that edge $\{b,h\}$ exists. Hence, the corresponding SA-E-noCC solution fulfills the SA constraint. For every OD delivery task $(b_{x_i},b_{x_j})$ with $i\neq j$ we consider the path $(b_{x_i},h_{x_i,y_i},h_{x_j,y_j},b_{x_j})$. As the $n$-Queens Completion Problem solution is feasible, $y_i\neq y_j$ and $|x_i-x_j|\neq |y_i-y_j|$. By construction, edge $\{h_{x_i,y_i},h_{x_j,y_j}\}$ with distance $1\le \phi$ exists. 
As $(x_i,y_i)$ is part of the $n$-Queens Completion Problem solution, there is no pre-placed queen $(x_i,j)$ with $y_i\neq j$ as they would attack each other. This gives, that edge $\{b_{x_i},h_{x_i,y_i}\}$ with distance $1\le \phi$ exists. By the same argument, edge $\{b_{x_j},h_{x_j,y_j}\}$ exists and the path is feasible.

Let a feasible SA-E-noCC solution be given. We show that the corresponding $n$-Queens Completion Problem solution is feasible by checking that there is a queen in each row and showing that none of them are attacking each other.
Assume, for contradiction, there is a row $x_i$ without a queen. 
Since a feasible SA-E-noCC solution exists, the OD delivery task
$(b_{x_i},b_{x_j})\in\mathcal W$, which exists by construction, must be
served by a path $(b_{x_i},h_{x_k,y_k},h_{x_l,y_l},b_{x_j})$.
The existence of the edge $\{b_{x_i},h_{x_k,y_k}\}$ implies $x_i=x_k$.
Thus, a queen is placed in row $x_i$, contradicting the assumption.
It remains to prove that the queens are pairwise non-attacking. Assume, for contradiction, that two queens placed on $(x_i,y_i)$ and $(x_j,y_j)$, where $i\neq j$, attack each other.
By construction, there is an OD delivery task $(b_{x_i},b_{x_j})$ with valid path $(b_{x_i},h_{x_i,y_i},h_{x_j,y_j},b_{x_j})$. As edge $\{h_{x_i,y_i},h_{x_j,y_j}\}$ exists, we have $x_i\neq x_j$, $y_i\neq y_j$ and $|x_i-x_j|\neq |y_i-y_j|$. This implies that those queens can not attack each other, contradicting the assumption.

\qed

\begin{lem}\label{lem6}
There exists a polynomial-time reductions from the branch to hub covering constraint (BH) problems to the edge-based covering constraint (E) problems:
\begin{itemize}
    \item SA-BH-CC reduces to SA-E-CC
    \item SA-BH-noCC reduces to SA-E-noCC
    \item MA-BH-CC reduces to MA-E-CC
    \item MA-BH-noCC reduces to MA-E-noCC
\end{itemize}
\end{lem}

\textbf{Proof:}

Recall that E variants require that \emph{every edge on a used path}
must have length at most $\phi$, whereas BH variants only require that \emph{every branch-to-hub edge on a used path} must have length at most $\phi$.

By adding a hub-to-hub edge of length $\phi$ between every pair of hubs, we ensure that the E constraint is always satisfied. Consequently, the E variants generalize the BH variants.





\qed

\begin{lem}\label{lem4}
There exists a polynomial-time reductions from the edge-based covering constraint (E) problems to the pseudo-metric total path covering constraint (TP) problems:
\begin{itemize}
    \item SA-E-CC reduces to pseudo-metric SA-TP-CC
    \item SA-E-noCC reduces to pseudo-metric SA-TP-noCC
    \item MA-E-CC reduces to pseudo-metric MA-TP-CC
    \item MA-E-noCC reduces to pseudo-metric MA-TP-noCC
\end{itemize}
\end{lem}

\textbf{Proof:}

Recall that E variants require that \emph{every edge on a used path}
must have length at most $\phi$, whereas TP variants additionally require that \emph{every 
path} must have length at most $\phi$.  
For a pseudo-metric graph, there exists a value 
$\alpha \in (0,1]$ such that, after dividing every hub-to-hub distance
by $\alpha$, the resulting graph is complete and metric.

As we will change $\phi$ and the distances in the reduction, we refer to the values in the E and TP setting by corresponding superscripts ($\phi^E$, $\phi^{TP}$, $d^E$ and $d^{TP}$).  
Given an E variant instance, we construct a corresponding TP instance as follows:

\begin{itemize}
    \item The set of branches, hubs, setup costs, OD delivery tasks, and the (optional) upper bound $m$ of open hubs remain unchanged.
    \item Add an edge between every pair of vertices, thereby making the graph complete.
    \item Set the distance of any branch-to-hub edge $\{b,h\}$ to $d_{bh}^{TP}=2$ iff $\{b,h\}$ existed in the E variant with $d_{bh}^E\leq \phi^E$. Otherwise, set that distance to $d_{bh}^{TP}=4$.
    \item Set the distance of any hub-to-hub edge $\{h,h'\}$ to $d_{hh'}^{TP}=1$ iff $\{h,h'\}$ existed in the E variant with $d_{hh'}^E\leq \phi^E$. Otherwise, set that distance to $d_{hh'}^{TP}=2$.
    \item Set the distance of any branch-to-branch edge $\{b,b'\}$ to $d_{bb'}^{TP}=4$.
    \item Set $\phi^{TP}=5$.
\end{itemize}

Figure \ref{figure7} illustrates a small E variant instance, while Figure \ref{figure8} shows the corresponding pseudo-metric TP instance obtained by this construction.

\begin{figure}[H]
    \centering

    \begin{minipage}[t]{0.40\textwidth}
        \centering
        \vspace{0pt}
        \begin{tikzpicture}[scale=0.7]
            
            \path[use as bounding box] (-0.6,1.5) rectangle (8.6,6.6);

            \node[shape=circle,draw=lightgray] (12) at (4,6) {$H_1$};
            \node[shape=circle,draw=lightgray] (22) at (4,4) {$H_2$};
            \node[shape=circle,draw=lightgray] (32) at (4,2) {$H_3$};

            \node[shape=circle,draw=blue,minimum size=20pt] (2) at (8,4) {$B_2$};
            \node[shape=circle,draw=blue,minimum size=20pt] (3) at (0,4) {$B_1$};

            \draw[-,line width=0.4mm] (12) edge node[left] {} (3);
            \draw[-,line width=0.4mm] (12) edge node[left] {} (2);
            \draw[-,line width=0.4mm] (22) edge node[left] {} (3);
            \draw[-,line width=0.4mm] (32) edge node[left] {} (3);
            \draw[-,line width=0.4mm] (22) edge node[left] {} (2);
            \draw[-,line width=0.4mm] (22) edge node[left] {} (32);

        \end{tikzpicture}

        \captionof{figure}{Example E variant instance, where black edges $e$ correspond to \newline $d_e^E\leq \phi^E$.}
        \label{figure7}
    \end{minipage}%
    \hspace{0.04\textwidth}%
    \begin{minipage}[t]{0.56\textwidth}
        \centering
        \vspace{0pt}
        \begin{tikzpicture}[scale=0.7]
            
            \path[use as bounding box] (-0.6,1.5) rectangle (13.8,6.6);

            \node[shape=circle,draw=lightgray] (12) at (4,6) {$H_1$};
            \node[shape=circle,draw=lightgray] (22) at (4,4) {$H_2$};
            \node[shape=circle,draw=lightgray] (32) at (4,2) {$H_3$};

            \node[shape=circle,draw=blue,minimum size=20pt] (2) at (8,4) {$B_2$};
            \node[shape=circle,draw=blue,minimum size=20pt] (3) at (0,4) {$B_1$};

            \draw[-, bend right, dashed,  line width=0.4mm] (12) edge node[left] {} (32);
            \draw[-, bend right=80, loosely dotted, color=black!50, line width=0.4mm] (2) edge node[left] {} (3);

            \draw[-,dashed,line width=0.4mm] (12) edge node[left] {} (3);
            \draw[-,dashed,line width=0.4mm] (12) edge node[left] {} (2);
            \draw[-,dashed,line width=0.4mm] (22) edge node[left] {} (3);
            \draw[-,dashed,line width=0.4mm] (32) edge node[left] {} (3);
            \draw[-,dashed,line width=0.4mm] (22) edge node[left] {} (2);
            \draw[-,color=black,line width=0.4mm] (22) edge node[left] {} (32);
            \draw[-,dashed,line width=0.4mm] (12) edge node[left] {} (22);
            \draw[-, loosely dotted, color=black!50, line width=0.4mm] (32) edge node[left] {} (2);

            \begin{scope}[shift={(10,3)}]
                \draw[black, line width=0.9pt] (0,2.5) -- (1.4,2.5);
                \node[anchor=west] at (1.8,2.5) {1};

                \draw[black,dashed,line width=0.9pt] (0,1.1) -- (1.4,1.1);
                \node[anchor=west] at (1.8,1.1) {2};

                \draw[black!50,loosely dotted,line width=1.1pt] (0,-0.3) -- (1.4,-0.3);
                \node[anchor=west] at (1.8,-0.3) {4};
            \end{scope}

        \end{tikzpicture}

        \captionof{figure}{Corresponding pseudo-metric TP instance constructed from the E variant, with distances $d_e^{TP}$ indicated in the legend.}
        \label{figure8}
    \end{minipage}

\end{figure}

The constructed graph is pseudo-metric, as for $\alpha=\frac{1}{2}$ it is a complete graph with distance $2$ or $4$ on every edge.
This transformation can be carried out in polynomial time.
In the following, we consider a solution for the E problem described by open hubs and paths for the OD delivery paths corresponding to a solution for the TP problem with the same open hubs and paths, and vice versa. To complete the reduction, we show that a feasible solution for the E problem corresponds to a feasible solution for the TP problem on the constructed instance with the same
objective function value and vice versa.

The objective function values are identical, as they only depend on the open hubs and setup costs, which remain
unchanged. 

For every feasible E variant solution with OD delivery task $(b,b')$, there is a solution path $(b,h,h',b')$ with $d_{bh}^E\leq \phi^E$, $d_{hh'}^E\leq \phi^E$ and $d_{h'b}^E\leq \phi^E$. As the OD delivery tasks and delivery paths remain unchanged, we need to check if that path is also feasible in the TP solution.
By construction, that path has distance $d_{bh}^{TP}+d_{hh'}^{TP}+d_{h'b'}^{TP}\leq   2+1+2=5\leq \phi^{TP}$. Hence, this path is also feasible in the TP solution.

Vice versa, every feasible TP variant solution path $(b,h,h',b')$ for the OD delivery task $(b,b')$ fulfills $d_{bh}^{TP}+d_{hh'}^{TP}+d_{h'b'}^{TP}\leq \phi^{TP}$. Assume, for contradiction, this path is invalid in the $E$ variant. Then either $d_{bh}^E> \phi^E$, $d_{hh'}^E> \phi^E$ or $d_{h'b'}^E> \phi^E$. We make a case distinction between the first and second cases, as the third is symmetrical to the first one.

If $d_{bh}^E> \phi^E$, by construction $d_{bh}^{TP}+d_{hh'}^{TP}+d_{h'b'}^{TP}\geq   4+0+2=6> \phi^{TP}$, contradicting the assumption that the TP solution is feasible.

If $d_{hh'}^E> \phi^E$, by construction $d_{bh}^{TP}+d_{hh'}^{TP}+d_{h'b'}^{TP}\geq   2+2+2=6> \phi^{TP}$, also contradicting the assumption that the TP solution is feasible.\qed

\begin{lem}\label{lem2}
There is a $|\mathcal{B}|^2$-approximation algorithm for MA-TP-noCC.
\end{lem}

\textbf{Proof:}
We present Algorithm \ref{alg:greed} and show afterwards, that it is a $|\mathcal{B}|^2$-approximation algorithm for MA-TP-noCC.
The algorithm solves each delivery task individually by opening the cheapest feasible hubs for each delivery task. This is done by first computing the costs for different routing options and afterwards picking the cheapest one for each delivery task.

\begin{algorithm}
\caption{Greedy algorithm for MA-TP-noCC}
\label{alg:greed}
\begin{algorithmic}[1]
\Require An MA-TP-noCC instance with branches $\mathcal{B}$, potential hubs $\mathcal{H}$,
         OD tasks $\mathcal{W}$, opening costs $c$, distances $d$, and threshold $\phi$
\Ensure A set of open hubs $S \subseteq \mathcal{H}$, or report infeasibility

\State $S \gets \emptyset$ \algmark{brace-x}

\vspace{2mm}
\ForAll{$b,b' \in \mathcal{B}$ and $h,h' \in \mathcal{H}$} \algmark{gamma-start}
    \If{$d_{bh} + d_{hh'} + d_{h' b'} > \phi$}
        \State $\gamma_{b,h,h',b'} \gets \infty$
    \ElsIf{$h \neq h'$}
        \State $\gamma_{b,h,h',b'} \gets c_h + c_{h'}$
    \Else
        \State $\gamma_{b,h,h',b'} \gets c_h$
    \EndIf
\EndFor \algmark{gamma-end}

\vspace{2mm}
\ForAll{$(b,b') \in \mathcal{W}$} \algmark{choice-start}
    \State Choose $(h^\star, h'^\star) \in
    \arg\min_{(h,h')\in \mathcal{H}\times\mathcal{H}}
    \gamma_{b,h,h',b'}$
    \If{$\gamma_{b,h^\star, h'^\star,b'} = \infty$}
        \State \Return no feasible solution
    \Else
        \State $S \gets S \cup \{h^\star, h'^\star\}$
    \EndIf
\EndFor \algmark{choice-end}

\State \Return $S$

\begin{tikzpicture}[remember picture,overlay]
    \coordinate (brace-xpos) at ($(brace-x.east)+(6.7cm,0)$);

    \coordinate (gamma-top) at (brace-xpos |- gamma-start.east);
    \coordinate (gamma-bottom) at (brace-xpos |- gamma-end.east);

    \draw[
        decorate,
        decoration={brace,amplitude=5pt},
        thick
    ]
    ($(gamma-top)+(0,0.8ex)$) --
    ($(gamma-bottom)+(0,0.4ex)$)
    node[midway,right=6pt,text width=3.2cm,align=left]
    {Define $\gamma$ as the opening cost of each valid path};

    \coordinate (choice-top) at (brace-xpos |- choice-start.east);
    \coordinate (choice-bottom) at (brace-xpos |- choice-end.east);

    \draw[
        decorate,
        decoration={brace,amplitude=5pt},
        thick
    ]
    ($(choice-top)+(0,0.8ex)$) --
    ($(choice-bottom)+(0,0.4ex)$)
    node[midway,right=6pt,text width=3.2cm,align=left]
    {Choose a cheapest valid path for each OD delivery task};
\end{tikzpicture}

\end{algorithmic}
\end{algorithm}

The algorithm runs in polynomial time, since it evaluates all tuples
$(b,h,h',b')\in \mathcal{B}\times\mathcal{H}\times\mathcal{H}\times\mathcal{B}$
and afterwards considers a subset as $\mathcal{W}\le |\mathcal{B}|^2$.

We first prove feasibility. For every OD delivery task, we consider path $(b,h^\star,h'^\star,b')$ with $h^\star,h'^\star$ are the hubs minimizing $\gamma_{b,h^\star,h'^\star,b'}$ as proposed in Step 12 of Algorithm \ref{alg:greed}. For every tuple $(b,h,h',b')$, the value
$\gamma_{b,h,h',b'}$ is finite if and only if the path
$(b,h,h',b')$ satisfies the TP covering constraint
\[
d_{bh}+d_{hh'}+d_{h' b'}\leq \phi .
\]
Thus, whenever the algorithm does not return infeasibility, it selects, for
each OD delivery task $(b,b')\in\mathcal{W}$, a pair of hubs
$(h^\star, h'^\star)$ that induces a feasible path
$(b,h^\star, h'^\star,b')$. Since both selected hubs are added to $S$,
the resulting hub set admits a feasible path for every OD delivery task.
Conversely, if the algorithm returns infeasibility for some
$(b,b')\in\mathcal{W}$, then
$\gamma_{b,h,h',b'}=\infty$ for all $h,h'\in\mathcal{H}$, and hence
no feasible path exists for this OD delivery task. Therefore, the instance is
infeasible.

It remains to bound the objective function value. For each OD delivery task
$(b,b')\in\mathcal{W}$, let $(h^\star_{bb'}, h'^\star_{bb'})$ be the hub
pair chosen by the algorithm. Since $S$ is the union of all selected hubs, we
have
\[
\sum_{h\in S} c_h
\leq
\sum_{(b,b')\in\mathcal{W}}
\gamma_{b,h^\star_{bb'},h'^\star_{bb'},b'} .
\]
Let $S^{\mathrm{OPT}}$ be an optimal hub set with objective function value
$\mathrm{OPT}=\sum_{h\in S^{\mathrm{OPT}}}c_h$. For every
$(b,b')\in\mathcal{W}$, the optimal solution contains some feasible path
$(b,h,h',b')$ with $h,h'\in S^{\mathrm{OPT}}$. Hence, by the choice of
$(h^\star_{bb'},h'^\star_{bb'})$ as a minimum-cost feasible hub pair,
\[
\gamma_{b,h^\star_{bb'},h'^\star_{bb'},b'}
\leq
\sum_{h\in S^{\mathrm{OPT}}} c_h
=
\mathrm{OPT}.
\]
Consequently,
\[
\sum_{h\in S} c_h
\leq
\sum_{(b,b')\in\mathcal{W}}
\gamma_{b,h^\star_{bb'},h'^\star_{bb'},b'}
\leq
|\mathcal{W}|\,\mathrm{OPT}
\leq
|\mathcal{B}|^2\,\mathrm{OPT}.
\]
Thus, the algorithm is a $|\mathcal{B}|^2$-approximation algorithm.\qed

We further established a $\tfrac{|\mathcal{H}|}{k}$-approximation algorithm for MA-TP-noCC with uniform setup costs, where $c_h=1$ for all $h\in \mathcal{H}$ and any constant $k$. 
Since this result is not required for our main approximation bounds, we defer the proof to Appendix~\ref{app:extra} as Proposition~\ref{lem:ma1b2}.

\section{Conclusion}\label{chap:c}

Hub covering problems (HCPs) extend the classical hub location framework by explicitly
incorporating distance-based covering constraints, which are essential in applications
such as urban logistics, telecommunication networks, cargo transportation, and
e-mobility infrastructures.

This paper establishes the first systematic approximation results for HCPs.
We showed that we can achieve the same approximation guarantees for SA-BH-noCC and MA-BH-noCC as for the Set Cover problem.
In contrast, we prove that SA-TP-noCC, SA-E-noCC, and all cardinality-constrained variants
do not admit any polynomial-time approximation guarantee, as deciding whether a feasible solution exists is NP-complete, unless $P=NP$.

A key aspect of our results is that these inapproximability bounds persist under strong structural assumptions such as metricity and completeness, while the approximation algorithms do not rely on them.

We establish polynomial-time reduction hierarchies:
\[
\text{MA-BH-noCC} \leq_p \text{MA-E-noCC} \leq_p \text{MA-TP-noCC}
\quad\text{and}\quad
\text{SA-BH-noCC} \leq_p \text{SA-E-noCC} \leq_p \text{SA-TP-noCC},
\]
which implies a monotonicity of algorithmic difficulty with respect to the covering constraint. 
For MA-TP-noCC and MA-E-noCC, we derived the first polynomial-time approximation algorithms.
Nevertheless, a substantial gap remains between the known lower bound of
$\Omega(\log|\mathcal{B}|)$ and the upper bound of $|\mathcal{B}|^2$
$\left( \text{or } \tfrac{|\mathcal{H}|}{k} \text{ in the uniform setup cost case}\right)$.
Closing this gap constitutes an important direction for future research.

Given the similarities between weighted Set Cover and certain HCP variants,
a natural next step is the adaptation of classical Set Cover heuristics,
such as the greedy algorithm of \cite{chvatal1979greedy},
to the HCP setting.
However, extending these techniques to MA-TP-noCC is nontrivial:
A feasible path may involve a pair of hubs, and selecting a particular hub pair
can reduce the cost of other routes due to shared hubs.
Promising heuristic directions include randomized strategies and MIP-based relaxations.
For instance, relaxing integrality constraints on hub opening variables and prioritizing hubs with large fractional values is a natural heuristic that warrants further investigation.

\subsection*{Declaration of generative AI and AI-assisted technologies in the
writing process}
During the preparation of this work the authors used ChatGPT in order
to improve language and readability. After using the tool, the authors reviewed and
edited the content as needed and take full responsibility for the content
of the publication.

\pagebreak
 \bibliographystyle{agsm}
 \bibliography{bib}  
 \pagebreak
\appendix

\section{Additional Approximation Result}\label{app:extra}

For the special case of uniform setup costs, we propose Algorithm \ref{alg:2}, which we bound afterwards.

\begin{algorithm}
\caption{$|\mathcal{H}|/k$-approximation algorithm for MA-TP-noCC with uniform setup costs}
\label{alg:2}
\begin{algorithmic}[1]
\Require An MA-TP-noCC instance with branches $\mathcal{B}$, potential hubs $\mathcal{H}$,
         OD tasks $\mathcal{W}$, distances $d$, threshold $\phi$, and fixed constant $k\in\mathbb{N}$
\Ensure A set of open hubs $S\subseteq\mathcal{H}$, or report infeasibility

\For{$i=1,\dots,k$} \algmark{smallsets-start}
    \ForAll{$S\subseteq \mathcal{H}$ with $|S|= i$}
        \State $\textsc{Feasible} \gets \textsc{true}$
        \ForAll{$(b,b')\in\mathcal{W}$}
            \If{there is no pair $(h,h')\in S\times S$ with
            $d_{bh}+d_{h h'}+d_{h' b'}\leq \phi$}
                \State $\textsc{Feasible} \gets \textsc{false}$
                \State \textbf{break}
            \EndIf
        \EndFor
        \If{$\textsc{Feasible}=\textsc{true}$}
            \State \Return $S$
        \EndIf
    \EndFor
\EndFor \algmark{smallsets-end}

\vspace{2mm}

\ForAll{$(b,b')\in\mathcal{W}$} \algmark{allhubs-start}
   \If{there is no pair $(h,h')\in \mathcal{H}\times\mathcal{H}$ with
$d_{bh}+d_{hh'}+d_{h' b'}\leq \phi$}
   
        \State \Return no feasible solution
    \EndIf
\EndFor \algmark{allhubs-end}
\State \Return $\mathcal{H}$

\begin{tikzpicture}[remember picture,overlay]
    \coordinate (brace-x) at ($(smallsets-start.east)+(9cm,0)$);

    \coordinate (smallsets-top) at (brace-x |- smallsets-start.east);
    \coordinate (smallsets-bottom) at (brace-x |- smallsets-end.east);

    \draw[
        decorate,
        decoration={brace,amplitude=5pt},
        thick
    ]
    ($(smallsets-top)+(0,0.8ex)$) --
    ($(smallsets-bottom)+(0,0.4ex)$)
    node[midway,right=6pt,text width=3.2cm,align=left]
    {Test hub sets of cardinality $i=1,2,...,k$};

    \coordinate (allhubs-top) at (brace-x |- allhubs-start.east);
    \coordinate (allhubs-bottom) at (brace-x |- allhubs-end.east);

    \draw[
        decorate,
        decoration={brace,amplitude=5pt},
        thick
    ]
    ($(allhubs-top)+(0,0.8ex)$) --
    ($(allhubs-bottom)+(0,0.4ex)$)
    node[midway,right=6pt,text width=3.2cm,align=left]
    {Test feasibility of opening all hubs};
\end{tikzpicture}

\end{algorithmic}
\end{algorithm}

\begin{lem}\label{lem:ma1b2}
For every fixed constant $k \in \mathbb{N}$, choosing a solution based on open hubs of Algorithm \ref{alg:2} yields a $\frac{|\mathcal{H}|}{k}$-approximation for MA-TP-noCC with uniform setup costs, where $c_h=1\ \forall h\in \mathcal{H}$.
\end{lem}

\textbf{Proof:}

We can construct paths for any OD delivery task by testing any pair of returned hubs and take an arbitrary feasible one.

For a fixed candidate set $S\subseteq \mathcal H$, the algorithm tests whether,
for every OD task $(b,b')\in\mathcal W$, there exists a hub pair
$(h,h')\in S\times S$ such that
\[
    d_{bh}+d_{hh'}+d_{h' b'}\leq \phi .
\]
This condition is equivalent to the existence of a feasible path
$(b,h,h',b')$ for every OD task using only hubs from $S$. Hence, whenever
the algorithm returns a candidate set $S$, each OD task can be served by a
feasible path using only hubs in $S$. Thus, the returned set is feasible.

If the algorithm does not find a feasible candidate set of cardinality at most
$k$, it performs the same feasibility test for $S=\mathcal H$. If this test is
successful, opening all hubs yields a feasible solution. If the test
fails, then there exists an OD task $(b,b')\in\mathcal W$ for which no pair
$(h,h')\in\mathcal H\times\mathcal H$ satisfies
\[
    d_{bh}+d_{hh'}+d_{h' b'}\leq \phi .
\]
Since every feasible solution can only use hubs from $\mathcal H$, no feasible
solution exists in this case. Therefore, the algorithm is correct.

For fixed $k$, the number of candidate sets considered in the enumeration phase
is bounded by
\[
    \sum_{i=1}^{k} \binom{|\mathcal H|}{i}
    =
    O(|\mathcal H|^k),
\]
which is polynomial in the input size for every fixed constant $k$. For each
candidate set $S$, the algorithm checks, for every OD task
$(b,b')\in\mathcal W$, all hub pairs in $S\times S$. Since $|S|\leq k$ during
the enumeration phase and $k$ is fixed, this takes polynomial time. The final
test for $S=\mathcal H$ checks all pairs in $\mathcal H\times\mathcal H$ for
each OD task, and hence also takes polynomial time. Thus, the overall running
time is polynomial for every fixed constant $k$.

Since the setup costs are uniform, the objective function value of a feasible hub set
$S$ is $|S|$. Let $\mathrm{OPT}$ denote the optimal objective function value, i.e., the
minimum number of hubs that must be opened in any feasible solution.

We distinguish two cases.

\smallskip
\noindent
\textbf{Case 1:} $\mathrm{OPT}\leq k$.  
Then there exists a feasible hub set of cardinality at most $k$. Since the
algorithm enumerates all hub sets of cardinality at most $k$, it will find such
a set before reaching the final test with $S=\mathcal H$. Moreover, because the
algorithm enumerates candidate sets in increasing cardinality, the first
feasible set it returns has minimum possible cardinality. Hence, the algorithm
returns an optimal solution.

\smallskip
\noindent
\textbf{Case 2:} $\mathrm{OPT}>k$.  
The algorithm tests opening $\mathcal{H}$, which is feasible as an optimal solution exists. Since $\mathrm{OPT}>k$, we have
\[
    |\mathcal H|
    \leq
    \frac{|\mathcal H|}{k}\,\mathrm{OPT}.
\]

In both cases, the algorithm returns a feasible solution of cost at most
\[
    \frac{|\mathcal H|}{k}\cdot \mathrm{OPT}.
\]
Therefore, for every fixed constant $k\in\mathbb N$, the algorithm is a
polynomial-time $\frac{|\mathcal H|}{k}$-approximation algorithm for
MA-TP-noCC with uniform setup costs.
\qed

\end{document}